\documentclass[aps, twocolumn, amsmath, showpacs, amssymb, prc, preprintnumbers]{revtex4-1}

\usepackage{graphicx}
\usepackage{dcolumn}
\usepackage{bm}

\newcommand{\be}{\begin{equation}}
\newcommand{\ee}{\end{equation}}
\newcommand{\ba}{\begin{eqnarray}}
\newcommand{\ea}{\end{eqnarray}}

\newcommand{\psat}{p_{\rm{sat}}}

\begin{document}

\title{Predictions for 5.023 TeV Pb+Pb collisions at the LHC} 

\author{H.~Niemi${}^{a}$,  K.~J.~Eskola${}^{b,c}$, R.~Paatelainen${}^{d}$, K.~Tuominen${}^{e,c}$}
\affiliation{$^a$Institut f\"ur Theoretische Physik, Johann Wolfgang Goethe-Universit\"at,
Max-von-Laue-Str. 1, D-60438 Frankfurt am Main, Germany}
\affiliation{$^{b}$University of Jyvaskyla, Department of Physics, P.O.~Box 35, FI-40014 University of Jyvaskyla, Finland}
\affiliation{$^c$Helsinki Institute of Physics, P.O.Box 64, FI-00014 University of Helsinki, Finland}
\affiliation{$^d$Departamento de Fisica de Particulas, Universidade de Santiago de Compostela,
E-15782 Santiago de Compostela, Galicia, Spain} 
\affiliation{$^e$Deparment of Physics, University of Helsinki, P.O.~Box 64, FI-00014 University of
Helsinki, Finland}

\begin{abstract}
We compute predictions for various low-transverse-momentum bulk observables in $\sqrt{s_{NN}} = 5.023$ TeV Pb+Pb collisions at the LHC from the event-by-event next-to-leading-order perturbative-QCD + saturation + viscous hydrodynamics ("EKRT") model. In particular, we consider the centrality dependence of charged hadron multiplicity, flow coefficients of the azimuth-angle asymmetries and correlations of event-plane angles. The centrality dependencies of the studied observables are predicted to be very similar to those at 2.76 TeV, and the magnitudes of the flow coefficients and event-plane angle correlations are predicted to be close to those at 2.76 TeV. The flow coefficients may, however, offer slightly more discriminating power on the temperature dependence of QCD matter viscosity than the 2.76 TeV measurements. Our prediction for the multiplicity in the 0-5\% centrality class, obtained using the two temperature-dependent shear-viscosity-to-entropy ratios that give the best overall fit to RHIC and LHC data is $dN_{\rm ch}/d\eta\big|_{|\eta|\le 0.5} =1876\dots2046$. We also predict a power-law increase from 200 GeV Au+Au collisions at RHIC to 2.76 and 5.023 TeV Pb+Pb collisions at the LHC, $dN_{\rm ch}/d\eta\big|_{|\eta|\le 0.5} \propto s^{0.164\dots0.174}$. 
\end{abstract}

\preprint{HIP-2015-37/TH}
\maketitle 

\section{Introduction}

The strategic goal of the ultrarelativistic heavy-ion physics programs at the Large Hadron Collider (LHC) and Relativistic Heavy Ion Collider (RHIC) is to understand the behavior of nuclear, hadronic and partonic matter under extreme conditions of temperature and density. These programs are designed to test the theory of the strong interaction, Quantum Chromodynamics (QCD), which predicts a new state of matter, the Quark-Gluon Plasma (QGP). To quantitatively understand the formation and transport properties of QCD matter, and QGP in particular, is the central driving force of the current LHC and RHIC heavy-ion experiments. This ambitious mission motivates also the new heavy-ion run at the LHC, where beams of lead ions are collided at a $\sqrt{s_{NN}} = 5.023$~TeV center-of-mass energy per nucleon-nucleon pair. 

To explore the QGP we have to reconstruct its properties from final-state observables measured in the heavy-ion experiments. The basic bulk observables for this purpose are hadronic multiplicities, transverse momentum ($p_T$) spectra, and Fourier harmonics ($v_n$) of azimuth-angle distributions. Measurements of these observables at LHC and RHIC provide convincing evidence for a formation of a strongly collective system and a nearly-thermalized low-viscosity QGP (for recent reviews, see \cite{Heinz:2013th,Niemi:2014lha}), whose expansion and cooling are describable with dissipative relativistic fluid dynamics \cite{Niemi:2015qia,Romatschke:2007mq,Luzum:2008cw,Schenke:2010rr,Gale:2012rq,Song:2010mg,Song:2011qa,Shen:2010uy,Song:2013qma,Bozek:2009dw,Bozek:2012qs, Niemi:2011ix, Niemi:2012ry, Karpenko:2015xea}. Additional observables, such as the measured event-by-event (EbyE) probability distributions of the relative $v_n$ fluctuations \cite{Aad:2013xma}, various event-plane angle correlations \cite{Aad:2014fla}, and also correlations of different flow harmonics \cite{ALICE:QM15} provide independent further constraints on the initial state and its fluctuations as well as the transport properties of QCD matter \cite{Niemi:2015qia}.

Consequently, it is of vital importance to understand the QCD dynamics of the formation and space-time evolution of the system produced in heavy-ion collisions. For this, one needs a comprehensive description of the different stages of heavy-ion collisions which bases as closely as possible on QCD -- and which preferably not only can reproduce the existing data but also predict observables for forthcoming heavy-ion runs. 

The details of initial isotropization and thermalization of QCD matter \cite{Kurkela:2015qoa} (for a recent review, see \cite{Kurkela:QM15}) remain still open. Nevertheless, the theoretical progress,
both in calculating the initial states from QCD \cite{Hirano:2005xf,Gale:2012rq,Schenke:2012wb,Drescher:2006pi,Drescher:2007ax,Eskola:1999fc,Eskola:2001bf,Eskola:2005ue,Paatelainen:2012at,Paatelainen:2013eea,Niemi:2015qia} and in describing the subsequent space-time evolution with dissipative fluid dynamics event by event, has recently brought us closer to the challenging goal of determining QCD matter properties such as shear viscosity \cite{Niemi:2015qia,Song:2010mg,Gale:2012rq,Bernhard:2015hxa}, bulk viscosity \cite{Noronha-Hostler:2014dqa, Ryu:2015vwa} and even equation of state \cite{Pratt:2015zsa} from the data. 
 
The key input for the heavy-ion fluid dynamics are the initial conditions, in particular the initial energy densities, formation times and flow conditions, which all fluctuate spatially as well as from one event to another. A viable QCD-based framework to compute such initial conditions is the next-to-leading order (NLO)-improved EKRT EbyE model introduced in detail in \cite{Niemi:2015qia}. This model describes remarkably consistently the centrality dependence of hadronic multiplicities, $p_T$ spectra and $v_n$ coefficients measured at mid-rapidity in $\sqrt{s_{NN}}=2.76$~TeV Pb+Pb collisions at the LHC and in 200~GeV Au+Au collisions at RHIC, as well as the fluctuation spectra of relative $v_n$'s and event plane angle correlations  measured at the LHC (see \cite{Niemi:2015qia}). As shown in \cite{Niemi:2015qia}, only with such a global analysis including both LHC and RHIC data one may systematically constrain the temperature dependence of the shear-viscosity-to-entropy ratio, $\eta/s(T)$, as a genuine matter property. In addition, owing to the fact that the EKRT initial states are obtained using NLO perturbative QCD (supplemented with a transparent saturation criterion), the NLO EbyE EKRT model has clear predictive power both in cms-energy and centrality -- which we will now exploit. 

In this paper, we employ the NLO-improved EKRT EbyE setup of Ref.~\cite{Niemi:2015qia} to compute predictions for the centrality dependence of charged hadron multiplicity and flow coefficients of the azimuth-angle asymmetries at mid-rapidity in $\sqrt{s_{NN}} = 5.023$ TeV Pb+Pb collisions at the LHC. In addition, we calculate also various correlations of two event-plane angles. To test the predictive power of this approach, we keep the few model parameters (explained below) fixed exactly as they were in \cite{Niemi:2015qia}. Especially, we will exploit the findings of \cite{Niemi:2015qia} that the best overall description of the LHC and RHIC data was obtained with $\eta/s(T)$ which ranges from a constant 0.2 to $\eta/s(T) = 0.12 + K(T/{\rm MeV}-150)$ with $K=-0.125/50$ for $T\le 150$~MeV and $K=0.525/350$ for $T\ge 150$~MeV. A particular goal here is to see whether any of the studied observables would show increased sensitivity to the QGP viscosity in these higher-energy collisions. 

\section{Details of the model}

We will next outline the NLO-improved EKRT EbyE model \cite{Niemi:2015qia}, and also set the stage for the initial state calculation performed in $\sqrt{s_{NN}} = 5.023$ TeV Pb+Pb collisions at the LHC. 

The initial minijet transverse energy $(E_T)$ produced perturbatively  into a rapidity interval $\Delta y$ in high-energy $A+A$ collisions and above a transverse momentum scale $p_0 \gg \Lambda_{\textrm{QCD}}$, can be computed in NLO as \cite{Paatelainen:2012at} 
\begin{equation}
\frac{dE_T}{d^2{\bf r}}(p_0, \sqrt{s_{NN}}, A, \Delta y, \mathbf{r}, \mathbf{b}; \beta) = \rho_{AA}(\mathbf{r})\sigma\langle E_T \rangle_{p_0,\Delta y,\beta},
\label{eq: dET}
\end{equation}
where $\mathbf{r}$ is the transverse coordinate and $\mathbf{b}$ impact parameter. The nuclear collision geometry is given by the nuclear overlap density 
\begin{equation}
\label{eq: ndensity}
\rho_{AA}(\mathbf{r}) = T_A \left (\mathbf{r} - \frac{{\bf b}}{2} \right )T_A \left (\mathbf{r} + \frac{{\bf b}}{2} \right ),
\end{equation}
where $T_A$ is the standard nuclear thickness function computed from the Woods-Saxon density distribution. The collinearly factorized minijet cross-section, $\sigma\langle E_T \rangle$ \cite{Eskola:1988yh,Eskola:2005ue}, is given in NLO by \cite{Eskola:2000ji, Eskola:2000my, Paatelainen:2012at}
\begin{equation}
\label{eq: sigmaet}
\sigma\langle E_T \rangle_{p_0,\Delta y,\beta} = \sum_{n=2}^{3}\frac{1}{n!}\int d[{\rm PS}]_n\frac{{\rm d}\sigma^{2\rightarrow n}}{d[{\rm PS}]_n} \tilde{S}_n({p_0,\Delta y,\beta}),
\end{equation}
where ${\rm d}\sigma^{2\rightarrow n}/d[{\rm PS}]_n$ are the differential partonic cross-sections, which contain the squared $2\rightarrow 3$ and ultraviolet renormalized $2\rightarrow 2$ matrix elements of order $\alpha_s^3$ \cite{Ellis:1985er,Paatelainen:2014fsa} and nuclear parton distribution functions (nPDFs). In this work, we use the NLO EPS09s spatially dependent nPDFs \cite{helenius:2012wd} together with the CTEQ6M free-proton PDFs \cite{Pumplin:2002vw}. The measurement functions $\tilde{S}_n$, which ensure that the NLO minijet $\sigma\langle E_T \rangle$ is a well-defined (collinear and infrared safe) quantity to compute, and which are analogous to those in jet production \cite{Kunszt:1992tn}, are given by \cite{Paatelainen:2012at}
\begin{equation}
\label{eq:mfunction}
\begin{split}
\tilde{S}_n({p_0,\Delta y,\beta}) &= \left ( \sum_{i=1}^n \Theta(y_i\in \Delta y)p_{T,i}\right ) \\
 &\times  \Theta \left (\sum_{i=1}^n p_{T,i}\geq 2p_0 \right ) \\
 &\times \Theta \left ( \sum_{i=1}^n \Theta(y_i\in \Delta y)p_{T,i} \geq \beta p_0 \right ),
\end{split}
\end{equation}
where $\Theta$ is the standard step function. Here, the individual terms on the r.h.s. of Eq. \eqref{eq:mfunction} define 
\textit{i)} the minijet $E_T$ as a sum of the transverse momenta $p_{T,i}$ of those partons whose rapidities $y_i$ fall in $\Delta y$, 
\textit{ii)} the hard perturbative scatterings to be those where at least a minimum amount $2p_0$ of transverse momentum is produced, and
\textit{iii)} a possible requirement of having a minimum amount of minijet transverse energy, $E_T\ge \beta p_0$, produced in $\Delta y$.
As explained in \cite{Paatelainen:2012at}, any $\beta \in [0,1]$ leads to a well-defined NLO computation but as the minijet $E_T$ is not a direct observable $\beta$ has to be determined from the data.  

In this model, the low-$p_T$ parton production is controlled by saturation of minijet $E_T$. As discussed in \cite{Paatelainen:2012at,Paatelainen:2013eea}, the saturation in $E_T$ production can be conjectured to take place when $(3\to 2)$ and higher-order partonic processes become of the same order of significance as the conventional $(2\to 2)$ processes,
\begin{equation}
\frac{{\rm d}E_T}{{\rm d}^2{\bf r}{\rm d}y}(2\to 2) \sim \frac{{\rm d}E_T}{{\rm d}^2{\bf r}{\rm d}y} (3 \to 2).
\end{equation}
This leads to the following local saturation criterion for the minijet $E_T$ production \footnote{This saturation condition is equivalent to that in the original LO EKRT model \cite{Eskola:1999fc} except that we now use the minijet $E_T$ and not the number of produced minijets.},
\begin{equation}
\label{eq: saturation}
\frac{{\rm d}E_T}{{\rm d}^2{\bf r}}(p_0,\sqrt{s_{NN}},A, \Delta y,{\bf r},{\bf b},\beta) = \left (\frac{K_{\rm{sat}}}{\pi} \right ) p_0^3 \Delta y,
\end{equation}
where the value of the parameter $K_{\rm{sat}}$ is determined from experimental data. For fixed $\beta$ and $K_{\rm{sat}}$, we can then solve Eq.~\eqref{eq: saturation} for the saturation scale, $p_0=p_{\rm sat}$, locally in $\mathbf{r}$.  

\begin{figure}[!h]
\includegraphics[width=8.5cm]{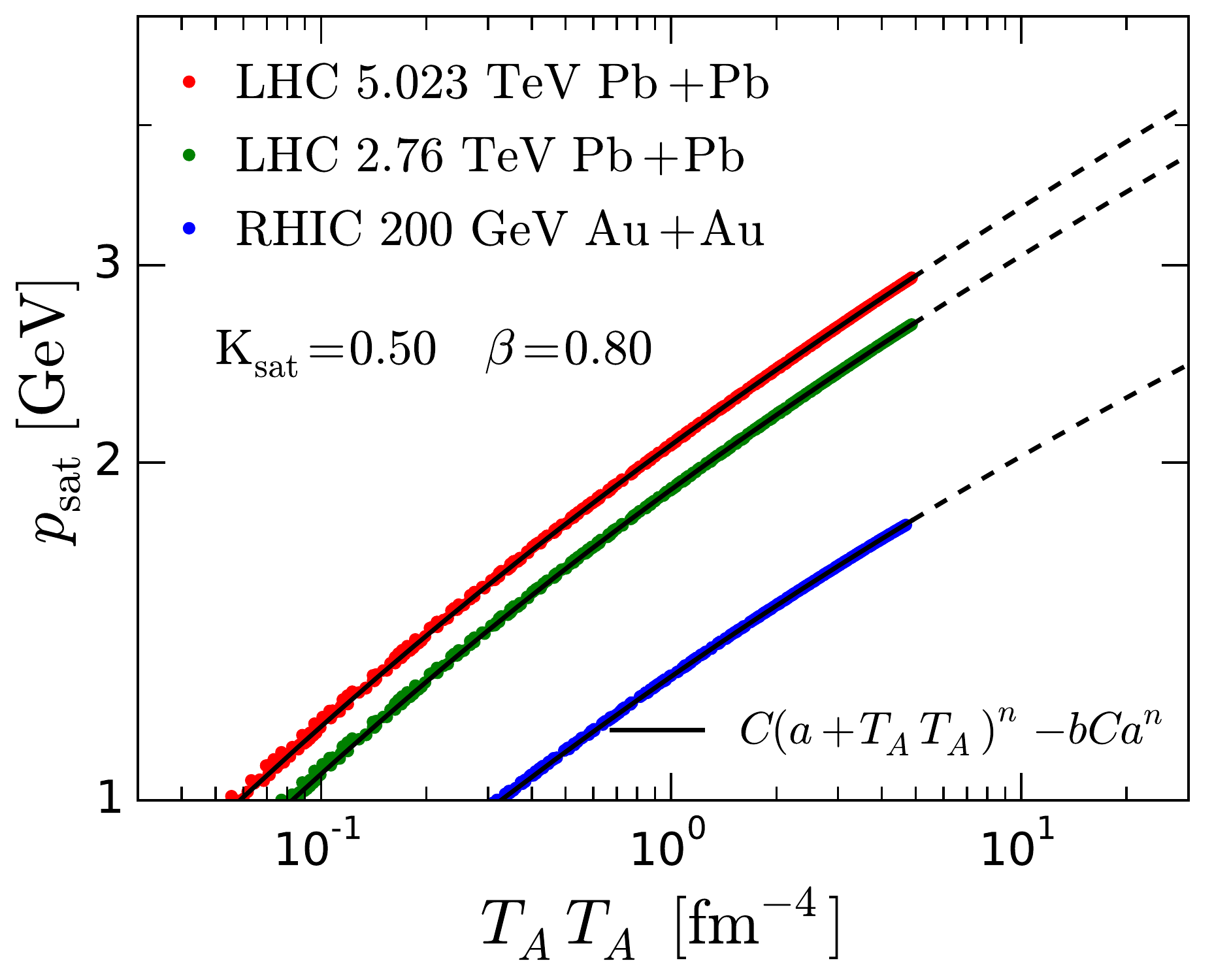}
\caption{\protect (Color online) The computed saturation momenta for fixed values of $K_{\rm sat} =0.5$ and $\beta =0.8 $ in LHC Pb+Pb collisions at $\sqrt{s_{NN}} = 5.023$ and 2.76 TeV and for RHIC Au+Au at $\sqrt{s_{NN}} = 200$ GeV, as functions of the nuclear overlap density $\rho_{AA}=T_AT_A$, for fixed impact parameters $b = 0$, $6.59$, and $8.27$ fm. The solid lines show the parametrization $p_{\rm sat}(\rho_{AA};K_{\rm sat},\beta)$ for each energy (see \cite{Niemi:2015qia} and Table \ref{tab:psat_parametrization_5023}). The dashed lines show extrapolations of these parametrizations into the high-$\rho_{AA}$ region outside the scope of the EPS09s nPDFs but which is occasionally probed in centralmost collisions. }
\label{fig:fitsaturation}
\end{figure}

In Fig.~\ref{fig:fitsaturation} we show, as one example, the computed $p_{\rm sat}$ values for $K_{\rm sat}=0.50$ and $\beta=0.8$ as a function of $\rho_{AA}$ with several values of impact parameter in $\sqrt{s_{NN}}=5.023$~TeV and 2.76~TeV Pb+Pb collisions at the LHC and in  $\sqrt{s_{NN}}=200$~GeV Au+Au collisions at RHIC. As demonstrated by the figure (and originally discussed already in the LO context \cite{Eskola:2001rx}) for fixed $\beta$ and  $K_{\rm sat}$  the computed $\psat$ scales quite accurately with $\rho_{AA}$. This key observation allows us to parametrize the computed $p_{\rm sat}(\rho_{AA})$ as a function of $K_{\rm sat}$ and $\beta$, and construct the EbyE fluctuating initial conditions for hydrodynamics. The parametrization, introduced in Ref.~\cite{Niemi:2015qia}, is given in Table \ref{tab:psat_parametrization_5023}. 

\begin{table}[h]
\caption{Parametrization of the saturation momentum $p_{\rm sat}(\rho_{AA}; K_{\rm sat}, \beta) = C\left[a + \rho_{AA}\right]^n - b C a^n$ where the ($K_{\rm sat}, \beta$) dependence of $a$, $b$, $C$ and $n$ is given by  $P_i(K_{\rm sat}, \beta) = a_{i0} + a_{i1} K_{\rm sat} + a_{i2} \beta + a_{i3}K_{\rm sat}\beta + a_{i4}\beta^2 + a_{i5} K_{\rm sat}^2$, in $\sqrt{s_{NN}} = 5.023$ TeV Pb+Pb collisions for $K_{\rm sat} \in [0.4, 2.0]$ and $\beta<0.9$.}
\begin{tabular}{c|cccc}
\hline
\hline
$P_i\rightarrow$ & $C$          & $n$                  & $a$             & $b$ \\  
\hline
$a_{i0}$	& 3.8815258    & 0.1473175           & -0.0033201     & 0.8542779 \\
$a_{i1}$	& -0.6898452   & -0.0192200          & 0.0146229      & -0.0780934 \\
$a_{i2}$	& 0.8721024    & -0.0341616          & -0.0006397     & 0.0945139 \\
$a_{i3}$	& 0.0514622    & -0.0016951          & 0.0090122      & -0.0018589 \\
$a_{i4}$	& -1.7354849   & 0.0597188           & -0.0024042     & -0.2031812 \\
$a_{i5}$	& 0.1329261    & 0.0059600           & -0.0020607     & 0.0288765 \\
\hline
\hline
\end{tabular}
\label{tab:psat_parametrization_5023}
\end{table}

To build up the EbyE initial conditions, we need to form the $\rho_{AA}$ locally in $\mathbf{r}$ for each nuclear collision event. This is done by first sampling the nucleon positions in the colliding nuclei from the standard Woods-Saxon density. Around each nucleon, we then set a gluon "cloud", a transverse density 
\begin{equation}
T_n(r) = \frac{1}{2\pi\sigma^2}\exp \left ( -\frac{r^2}{2\sigma^2}\right )
\end{equation}
and compute the thickness functions $T_A$ locally in each event as a sum of the corresponding gluon transverse densities at each transverse point. The width parameter $\sigma = 0.43$ fm is obtained from the exclusive electroproduction data of $J/\Psi$ collected by ZEUS at HERA \cite{Chekanov:2004mw}. 

For fixed $K_{\rm sat},\beta$, the local saturation scales $p_{\rm sat}(\mathbf{r}) = p_{\rm sat}(\rho_{AA}(\mathbf{r});K_{\rm sat},\beta)$ in each event are next obtained from the parametrization of Table~\ref{tab:psat_parametrization_5023}. The local EbyE fluctuating initial energy densities at the formation of the system can then be computed as 
\begin{equation}
\varepsilon(\mathbf{r}, \tau_s(\mathbf{r}) ) = \frac{dE_T(p_{\rm sat})}{d^2{\bf r}} \frac{1}{\tau_s(\mathbf{r})\Delta y} = \frac{K_{\rm{sat}}}{\pi}[p_{\rm sat}(\mathbf{r})]^4,
\label{eq:edensity}
\end{equation}
where the local formation time of the minijet plasma is given by $\tau_s(\mathbf{r}) = 1/p_{\rm sat}(\mathbf{r})$. Here, for the applicability of pQCD, the minimum allowed saturation scale is fixed to $p_{\rm sat}^{\rm min} = 1$~GeV. This corresponds to a formation time $\tau_0=0.2$~fm which we also take as the starting time of the viscous hydrodynamics stage. The evolution from $\tau_s({\bf r})$ to $\tau_{\rm 0}$ is here done simply by using 1 D Bjorken hydrodynamics (see discussion in \cite{Paatelainen:2013eea}).  

\begin{figure}[!h]
\includegraphics[width=8.7cm]{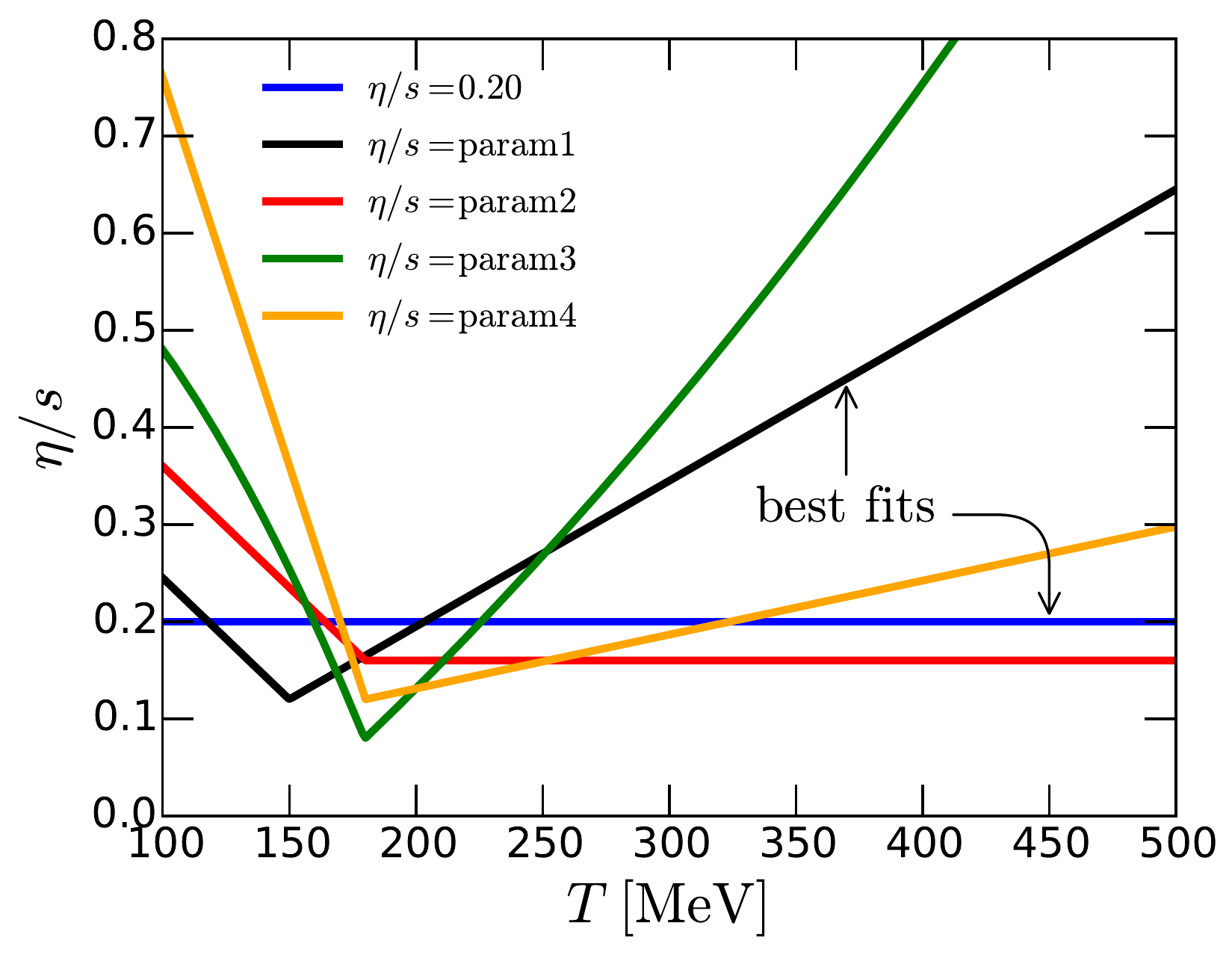}
\caption{\protect (Color online) Shear viscosity-to-entropy ratio as a function of temperature as determined in Ref.~\cite{Niemi:2015qia} and applied here. Out of these, the best overall global fit to the RHIC and LHC observables is obtained with $\eta/s=0.2$ and $param1$ \cite{Niemi:2015qia}.}
\label{fig:etapers}
\end{figure}

The fluid-dynamical evolution in each event is described by the state-of-the-art 2+1 dimensional dissipative relativistic hydrodynamic simulation previously employed in \cite{Niemi:2015qia}.  In this setup, we neglect the effects of bulk viscous pressure and diffusion currents, and the evolution equation of the shear-stress tensor $\pi^{\mu\nu}$ is given by transient fluid dynamics~\cite{IS,Denicol:2012cn,Molnar:2013lta}. The initial $\pi^{\mu\nu}(\tau_0)$ and transverse flow $v_T(\tau_0)$ are set to zero.  Exactly as in \cite{Niemi:2015qia}, we model the temperature dependence of $\eta/s(T)$ with the parametrizations shown in Fig.~\ref{fig:etapers}. 
As discussed in detail in \cite{Niemi:2015qia}, all of these reproduce the centrality dependence of charged hadron multiplicities and $p_T$ spectra at the LHC and RHIC, and $v_n$ flow harmonics at the LHC. However, two of them, $\eta/s=0.2$ and $param1$ do the best job in predicting simultaneously also the $v_n$ coefficients at RHIC and event-plane angle correlations at the LHC --- hence, a special emphasis is given to these two parametrizations in the predictions we present here. 

In the fluid-dynamic evolution, we employ the lattice QCD and hadron resonance gas based equation of state $s95p$-PCE-v1~\cite{Huovinen:2009yb} with a chemical freeze-out temperature $175$ MeV and kinetic freeze-out temperature $100$ MeV. The hadron spectra are calculated with the Cooper-Frye freeze-out procedure \cite{Cooper:1974mv}, and resonance decays after the freeze-out are included. The local equilibrium particle distributions are derived by using the Israel's and Stewart's 14-moment approximation \cite{Israel:1979wp}, where the relative deviations from the equilibrium distributions (the "$\delta f$ corrections") are assumed to be proportional to $p_\mu p_\nu \pi^{\mu\nu}$ for each particle species.   

As discussed in \cite{Paatelainen:2012at,Niemi:2015qia} the parameters $K_{\rm sat}$ and $\beta$ are correlated. Following again Ref.~\cite{Niemi:2015qia}, we fix $\beta=0.8$ and for each of the $\eta/s(T)$ parametrizations we iteratively determine $K_{\rm sat}$ so that we reproduce the charged hadron multiplicity measured by ALICE \cite{Aamodt:2010cz} in the 0-5 \% centrality class in $\sqrt{s_{NN}}=2.76$~TeV Pb+Pb collisions at the LHC. With the fixed $K_{\rm sat}$ and $\beta$ the NLO EKRT EbyE model is then closed, and it correctly predicts the centrality dependence of the LHC bulk observables and that for RHIC as well \cite{Niemi:2015qia}. Here, we wish to test and demonstrate this predictive power further, by computing bulk observables for the forthcoming higher-energy LHC heavy-ion run.

\section{Results}

\begin{figure*}[!]
\hspace{-0.5cm} 
\includegraphics[width=9.0cm]{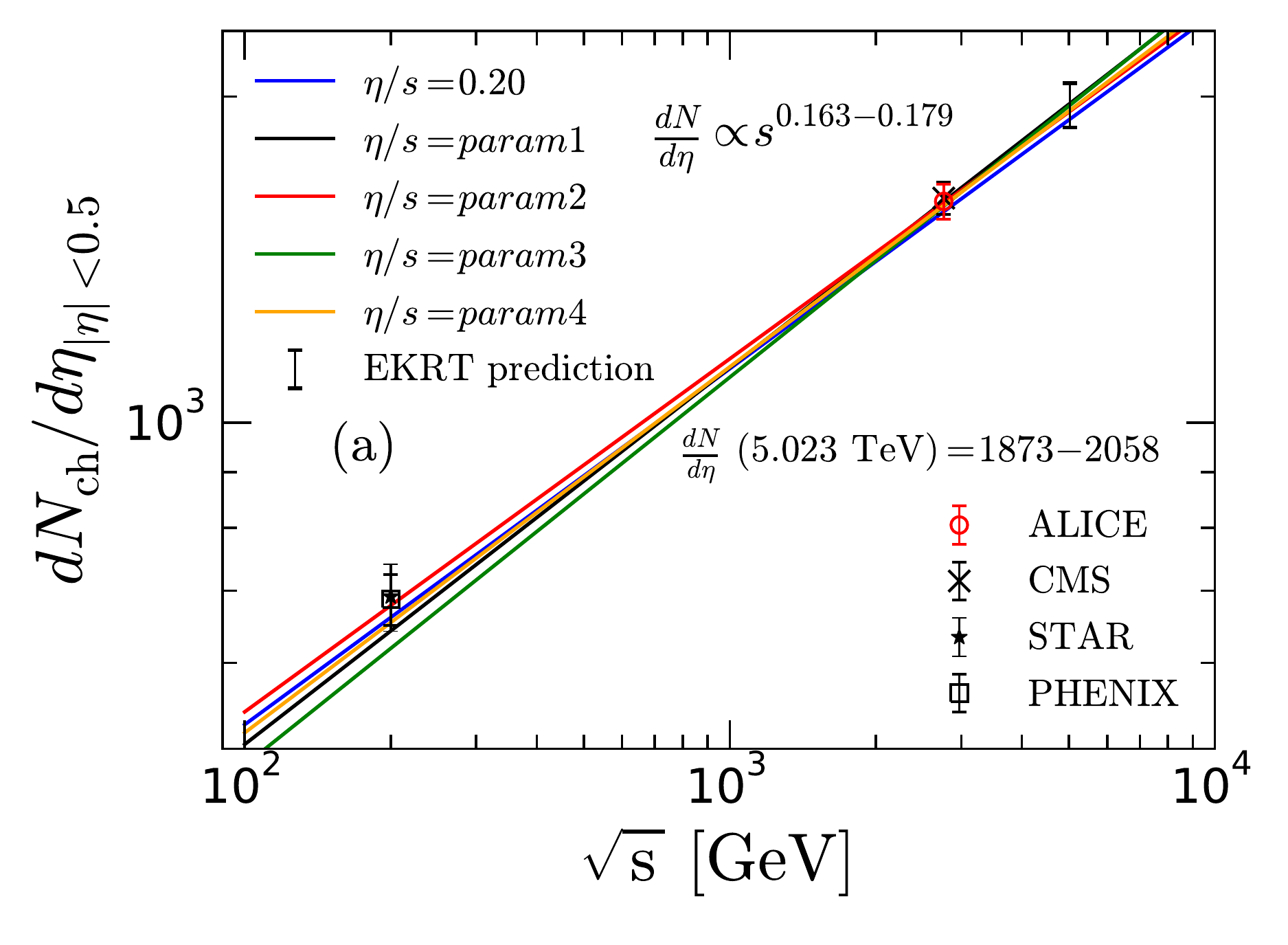}
\includegraphics[width=9.0cm]{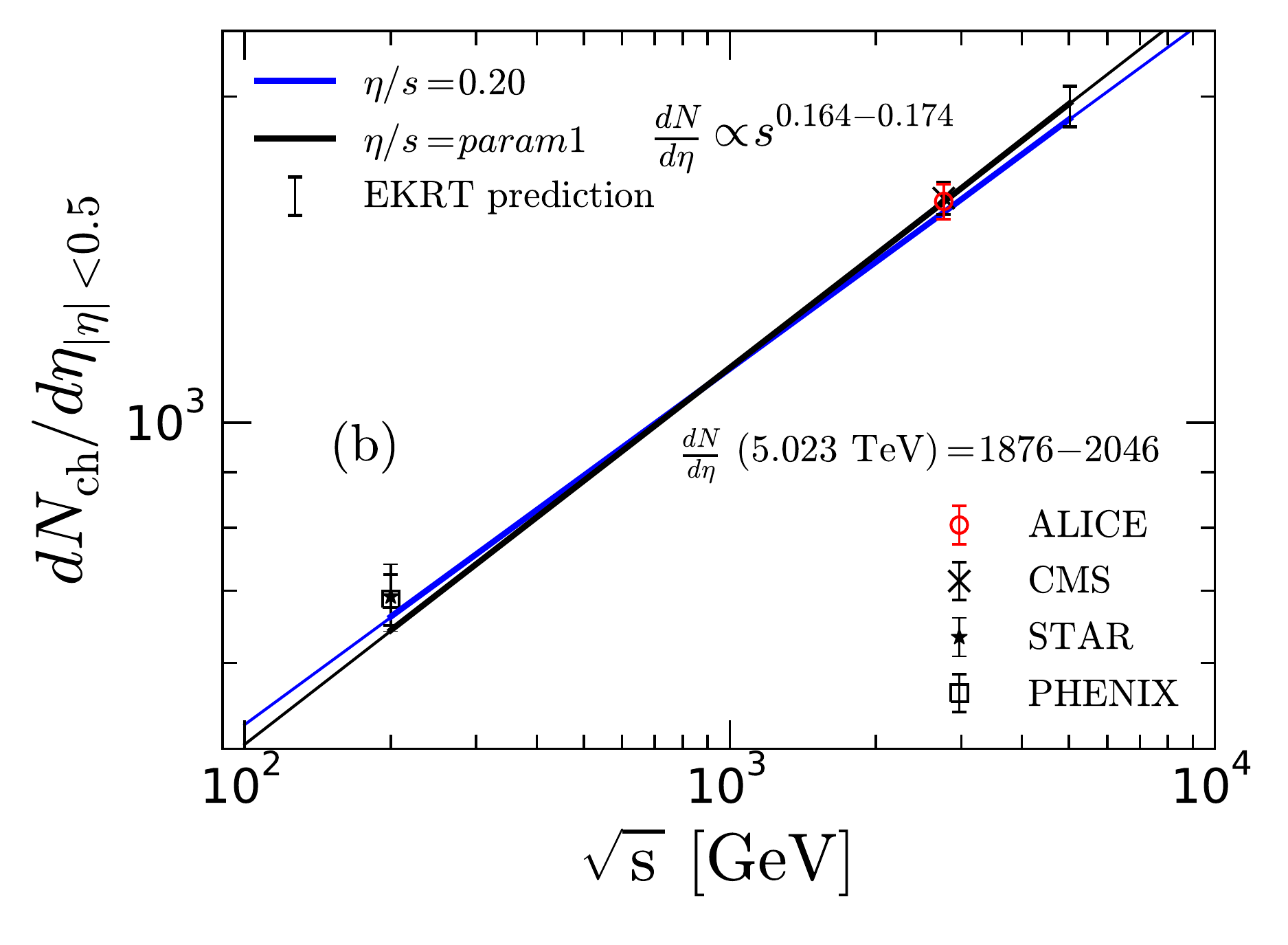}
\caption{\protect (Color online) Charged hadron multiplicities in the 0-5 \% centrality class in $\sqrt{s_{NN}} = 5.023$ TeV and 2.76 TeV Pb+Pb collisions at the LHC and 200 GeV Au+Au collisions at RHIC, computed (a) for all five $\eta/s(T)$ parametrizations of Fig.~\ref{fig:etapers} \cite{Niemi:2015qia} and (b) for the two best-fitting $\eta/s(T)$ parametrizations. The solid lines are power law fits to the EKRT results, see Table~\ref{tab:mult}. For explanation of the the EKRT error bars, see text. The ALICE \cite{Aamodt:2010cz} data point is used for normalization here, and the CMS multiplicity \cite{Chatrchyan:2011pb} is shown for comparison. The RHIC data are from STAR \cite{Abelev:2008ab} and PHENIX \cite{Adler:2004zn}. }
\label{fig:multiplicity}
\end{figure*}

In Fig.~\ref{fig:multiplicity}a and Table \ref{tab:mult} we present the NLO EKRT EbyE model predictions for the cms-energy dependence of charged hadron multiplicity in the 0-5 \% centrality class from RHIC $\sqrt{s_{NN}} = 200$~GeV Au+Au collisions to the LHC $\sqrt{s_{NN}} = 2.76$ and  5.023 TeV Pb+Pb collisions, computed for the five $\eta/s(T)$ parametrizations in Fig.~\ref{fig:etapers}. The three multiplicities in each case follow very closely the power law-fits of Table \ref{tab:mult} and shown by the lines in Fig.~\ref{fig:multiplicity}a. The computed RHIC and lower-energy LHC results are from \cite{Niemi:2015qia} and the predictions for the 5.023 TeV can be read off from where the EKRT error bar resides \footnote{As seen in the figure, the accuracy of our iteration in fixing $K_{\rm sat}$ in \cite{Niemi:2015qia} was a few percent, hence the curves do not all cross exactly in the middle of the 2.76 TeV ALICE point.}. Interestingly, the $\sqrt{s_{NN} }$ slopes get ordered according to the $\eta/s$ in the hottest QGP stages, $param3$ giving the steepest and $param2$ the gentlest slope. To estimate the error bar shown on our EKRT prediction, we allow for a change in our normalization  ($K_{\rm sat}$ from the LHC 2.76 TeV measurements) within the error limits of the ALICE multiplicity, and assume that a small change of $K_{\rm sat}$ does not affect the $\sqrt{s_{NN}}$ slope of multiplicity visibly. The EKRT error bar of Fig.~\ref{fig:multiplicity}a is then obtained as the envelope of the minimum and maximum predictions from all five $\eta/s$ parametrizations.

\begin{table}[h]
\caption{Charged hadron multiplicities in the 0-5\% centrality class in $\sqrt{s_{NN}} = 200$~GeV Au+Au, and 2.76 TeV and 5.023 TeV Pb+Pb collisions, computed from the NLO EbyE EKRT model for the five $\eta/s(T)$ cases of Ref.~\cite{Niemi:2015qia}, normalizing the $K_{\rm sat}$ to the 2.76 TeV measurement and keeping $\beta=0.8$ fixed. The last column shows a power-law fit to the computed three points.}

\begin{tabular}{ccccc}
\hline
\hline
\\
$\sqrt{s_{NN}}$/GeV &  200~~ & 2760~~ & 5023~~ & fit $a(s_{NN}/{\rm GeV^2})^p$ \\~\\
\hline
$\eta/s(T)=0.2$     & 662.1    & 1564    & 1908     & $a=115.9,  \, p=0.1643$ \\
param1              & 643.2    & 1599    & 1973     & $a=101.7,  \, p=0.1739$ \\
param2              & 679.6    & 1591    & 1941     & $a=120.6,  \, p=0.1630$ \\
param3              & 619.7    & 1583    & 1965     & $a=92.79,  \, p=0.1791$ \\
param4              & 655.6    & 1583    & 1945     & $a=109.2,  \, p=0.1689$ \\
\hline
\hline
\end{tabular}
\label{tab:mult}
\end{table}

Our best shot at the LHC 5.023 TeV multiplicity prediction is then presented in Fig.~\ref{fig:multiplicity}b where we focus only on the two best-fitting $\eta/s$ cases of Ref.~\cite{Niemi:2015qia}, the constant 0.2 and $param1$. We can see here and also in Fig.~\ref{fig:multiplicity}a that the variation in $\eta/s$ (within the five $\eta/s$ limits studied here) induces only a few-percent error on our LHC multiplicity prediction. Again, we estimate the EKRT error band similarly as for Fig.~\ref{fig:multiplicity}a, allowing our normalization to change within the limits of the 2.76 TeV ALICE point for both $\eta/s$ parametrizations and then taking a simple envelope of the maximum and minimum predictions. Hence, our prediction for the charged hadron multiplicity in the 0-5\% centrality class of 5.023 TeV Pb+Pb collisions at the LHC is 
\begin{equation}
\frac{dN_{\rm ch}}{d\eta}\bigg|_{|\eta|\le 0.5} =1876\dots2046.
\end{equation} 
The power law behavior the NLO EKRT model predicts here from 200 GeV Au+Au collisions to 5.023 TeV Pb+Pb collisions is 
\begin{equation}
\frac{dN_{\rm ch}}{d\eta}\bigg|_{|\eta|\le 0.5} \propto s^{0.164\dots0.174},
\end{equation}
with the smaller power for $param1$ and larger one for $\eta/s=0.2$.

\begin{figure}[!h]
\includegraphics[width=9.0cm]{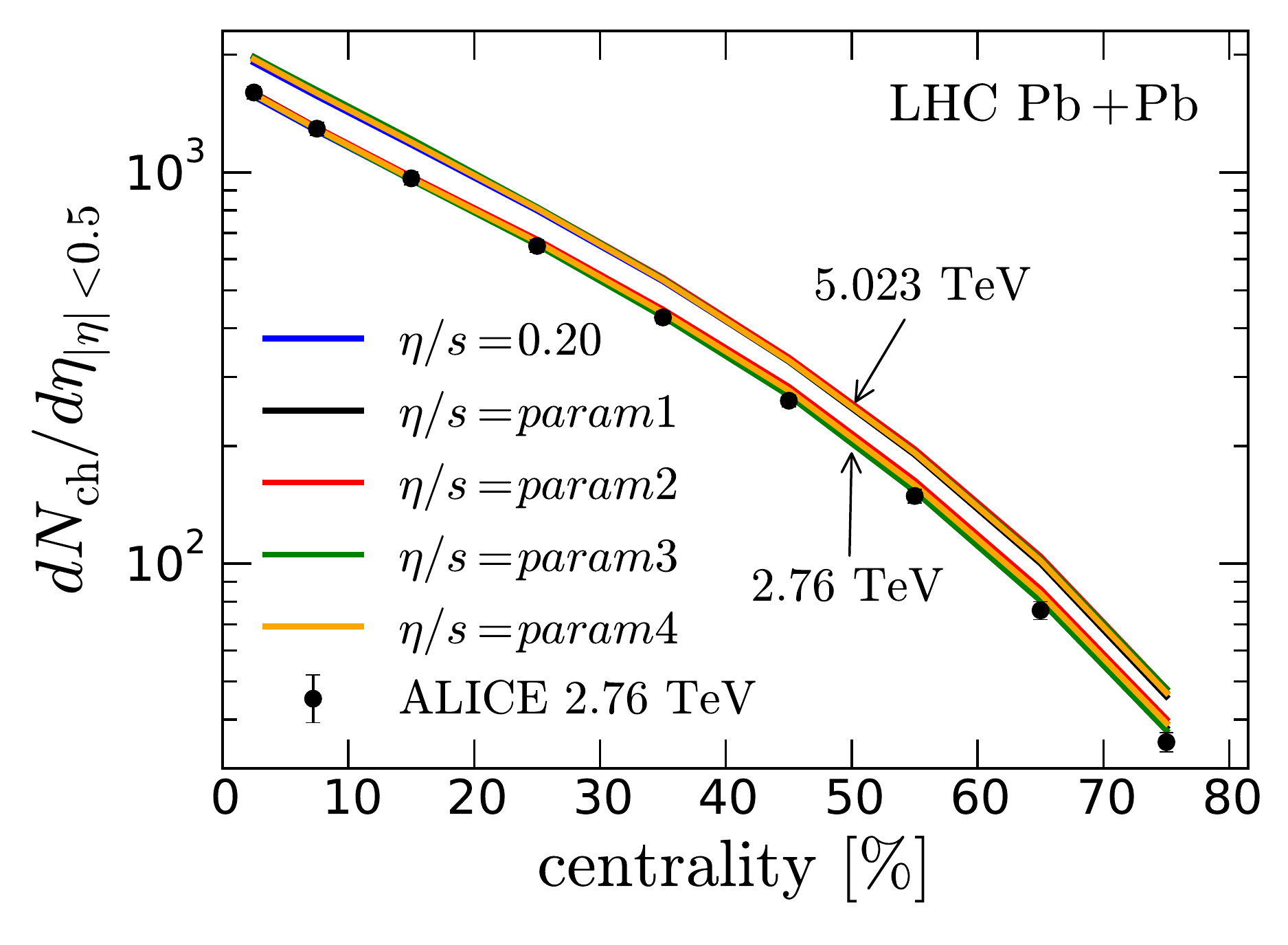}
\caption{\protect (Color online) 
The EKRT prediction for the centrality dependence of charged hadron multiplicity in $\sqrt{s_{NN}} = 5.023$ TeV Pb+Pb collisions at the LHC, computed for for all five $\eta/s(T)$ parametrizations of Fig.~\ref{fig:etapers} \cite{Niemi:2015qia}. The results of Ref.~\cite{Niemi:2015qia} and the ALICE measurement \cite{Aamodt:2010cz} at 2.76 TeV are shown for comparison. }
\label{fig:multcent}
\end{figure}

Figure \ref{fig:multcent} shows the NLO EKRT EbyE model prediction for the centrality dependence of the charged hadron multiplicity at mid-rapidity in 5.023 TeV Pb+Pb collisions. As seen in the figure, the computed centrality dependence remains very similar to that at 2.76 TeV, all five $\eta/s(T)$ cases leading essentially to the same result.  

\begin{figure*}[!]
\includegraphics[width=8.5cm]{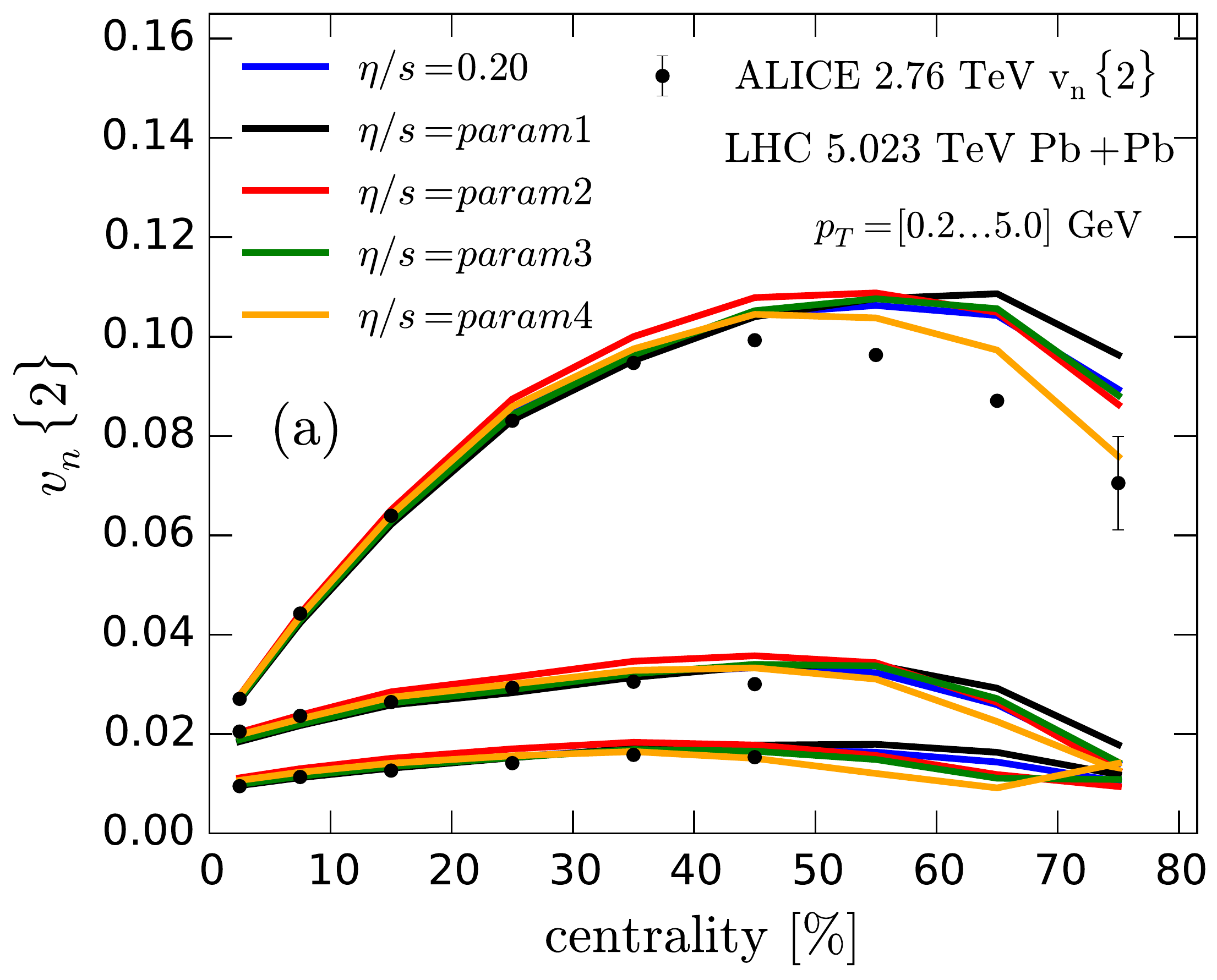}
\includegraphics[width=8.5cm]{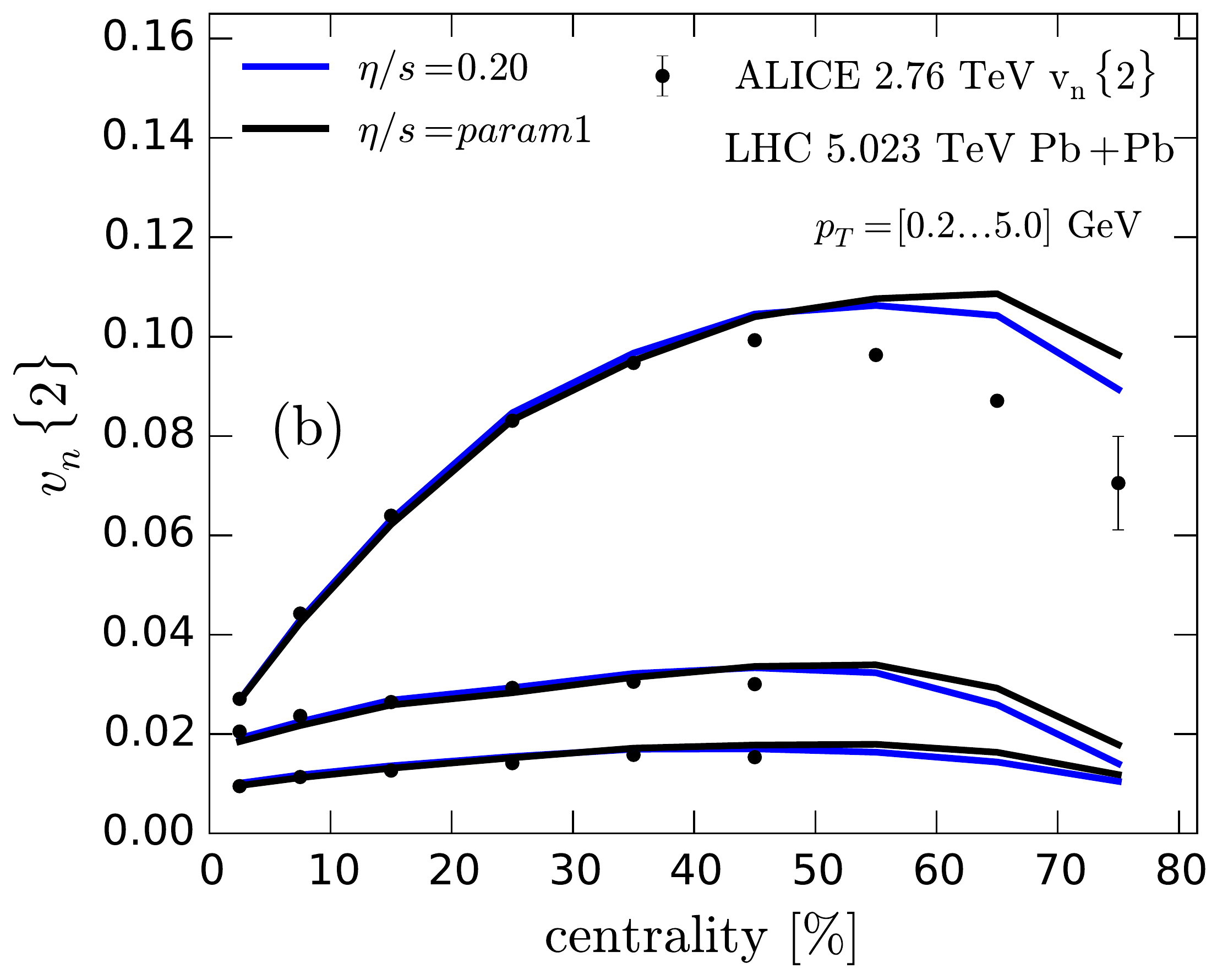}
\caption{\protect (Color online) Centrality dependence of the flow coefficients $v_n\{2\}$ from the charged hadron 2-particle cumulants in $\sqrt{s_{NN}} = 5.023$ TeV Pb+Pb collisions at the LHC, computed for (a) all five $\eta/s(T)$ parametrizations of Fig.~\ref{fig:etapers} \cite{Niemi:2015qia}, and (b) focusing on the two best fitting $\eta/s(T)$ scenarios. ALICE data \cite{ALICE:2011ab} at 2.76 TeV are shown for comparison.}
\label{fig:flowcoef}
\end{figure*}

\begin{figure*}[!]
\includegraphics[width=16.5cm]{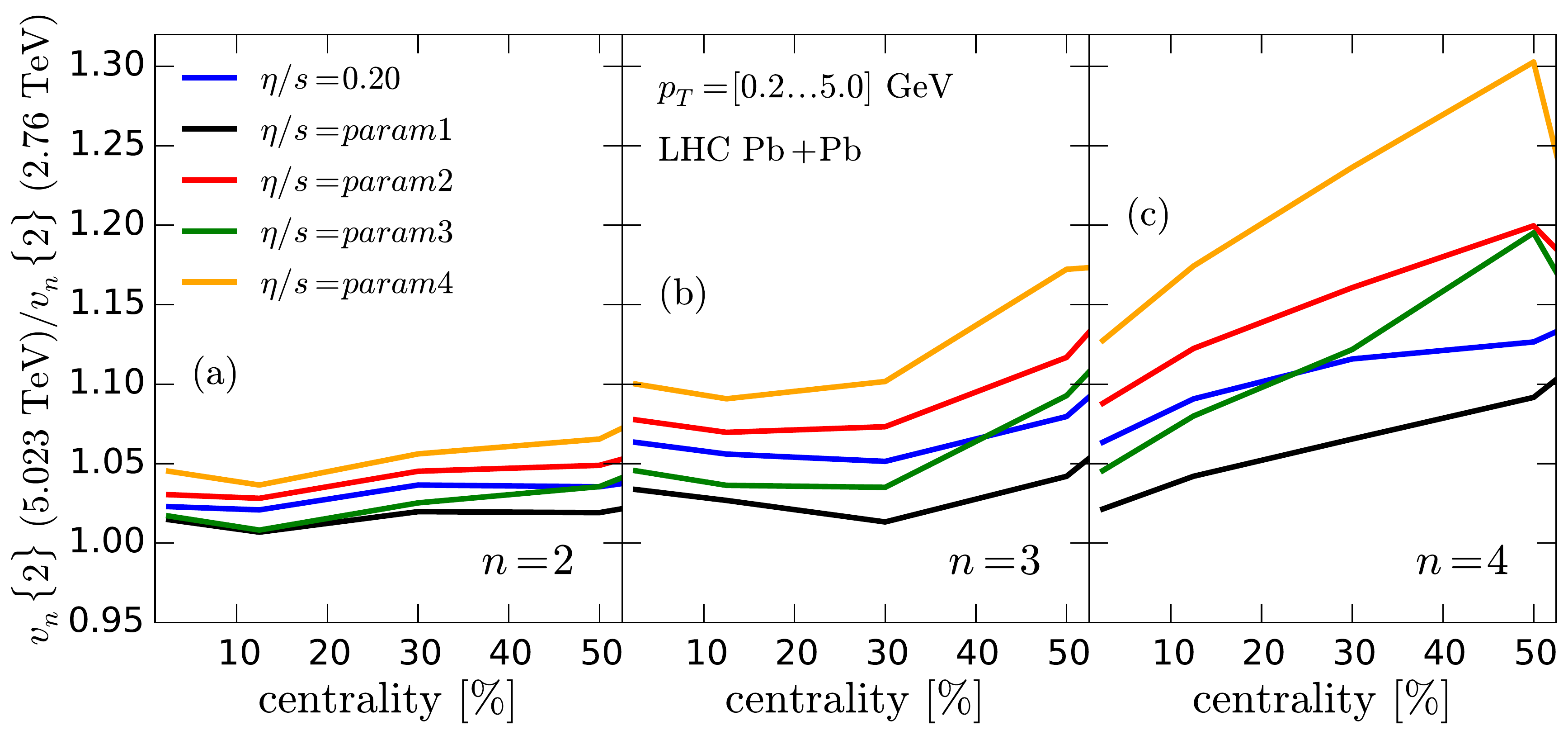}
\caption{\protect (Color online) Ratio of the flow coefficients $v_n\{2\}$ at 5.023 TeV and 2.76 TeV Pb+Pb collisions at the LHC for the five $\eta/s(T)$ parametrizations of Fig.~\ref{fig:etapers}.}
\label{fig:flowcoef_ratio}
\end{figure*}

In Fig.~\ref{fig:flowcoef} we show the EKRT 5.023 TeV predictions for the centrality dependence of flow harmonics $v_n\{2\}$ of charged hadrons obtained from 2-particle cumulants. ALICE 2.76 TeV data are shown for comparison here. Panel (a) shows our predictions for all the five $\eta/s(T)$ parametrizations studied, and panel (b) shows our best predictions. Interestingly, when comparing with Fig.~14a of Ref.~\cite{Niemi:2015qia}, we notice that when including the 5.023 TeV collisions into the analysis the $v_n\{2\}$ coefficients show slightly more sensitivity to $\eta/s(T)$ than using the 2.76 TeV data alone. This is further demonstrated in Fig.~\ref{fig:flowcoef_ratio}, where the ratios of the $v_n\{2\}$ coefficients at 5.023 TeV and 2.76 TeV are shown for the five different $\eta/s(T)$ parametrizations we have considered. Note that while the flow coefficients are more sensitive to the high-temperature part of $\eta/s(T)$ at higher collision energy, they still remain sensitive to the minimum value in the QCD transition region. Even if our best predictions $\eta/s = 0.20$ and $param1$ have quite different high temperature behavior, they also have different minimum values, and therefore both parametrizations give quite a similar increase of $v_2\{2\}$. Overall, the increases of $v_2\{2\}$ from 2.76 to 5.023 TeV are quite moderate, of the order 5 \% at most from central to mid-peripheral collisions. The increases of the higher harmonics $v_3\{2\}$ and $v_4\{2\}$ are more pronounced, and can be of the order 10 \% for the best $\eta/s(T)$ parametrizations, and go up to 30 \% if all the parametrizations are considered. The higher harmonics are also more sensitive to the differences between the $\eta/s(T)$ parametrizations than $v_2\{2\}$. Therefore, precise measurements of $v_n\{2\}$ should be able to give further constraints to $\eta/s(T)$.


\begin{figure*}
\includegraphics[width=\textwidth]{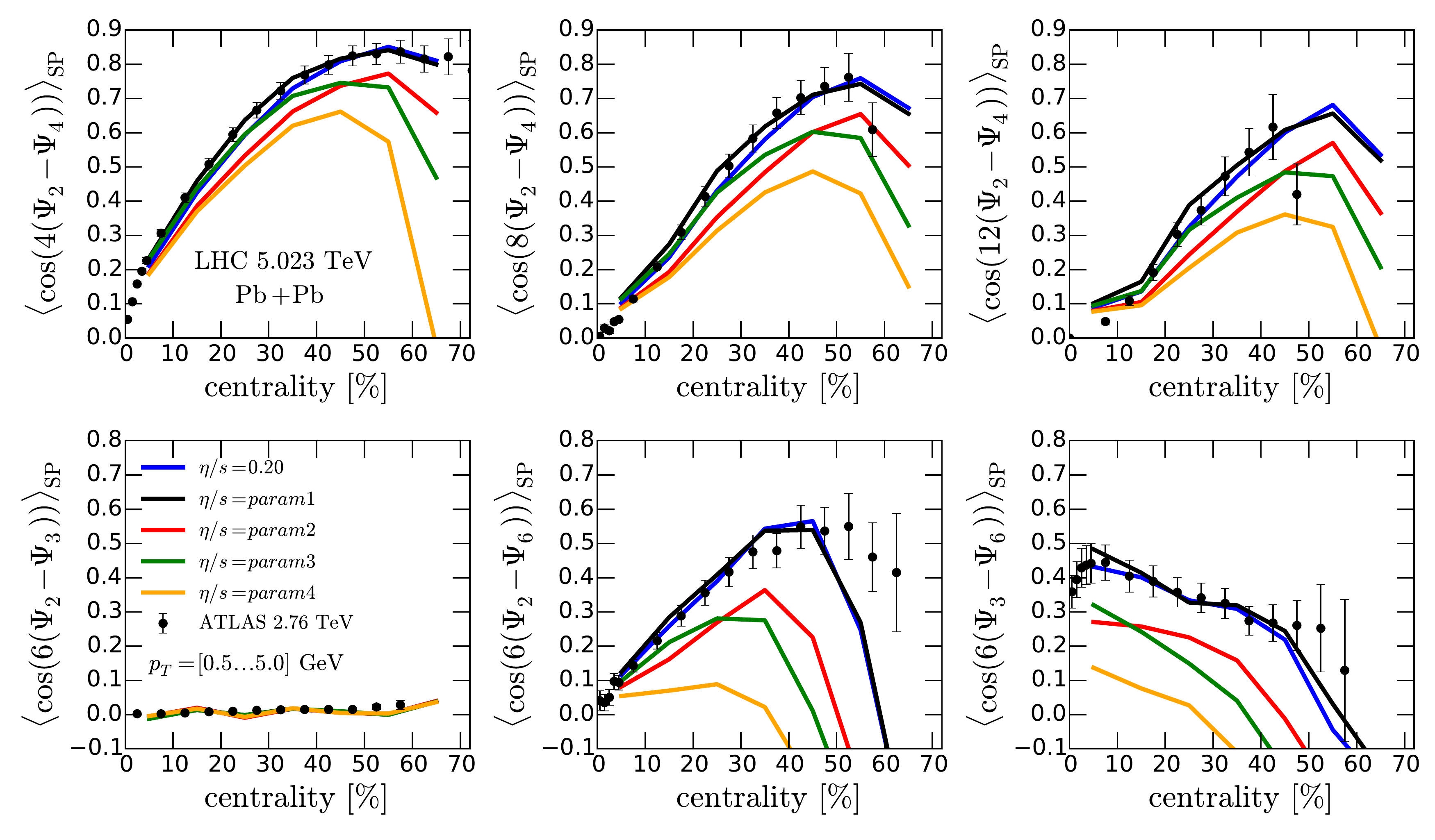}
\caption{\protect (Color online) Correlations of two event-plane angles for charged hadrons in $\sqrt{s_{NN}} = 5.023$ TeV Pb+Pb collisions at the LHC, computed from the NLO EKRT EbyE model using all five $\eta/s(T)$ cases of Fig.~\ref{fig:etapers} \cite{Niemi:2015qia}. The ATLAS 2.76 TeV data \cite{Aad:2014fla} are shown for comparison.}
\label{fig:EPcorrelations}
\end{figure*}

Finally, in  Fig.~\ref{fig:EPcorrelations} we show the NLO EKRT EbyE model predictions for several event-plane angle correlations involving two different event plane angles $\Psi_n$ of charged hadrons produced in $\sqrt{s_{NN}} = 5.023$ TeV Pb+Pb collisions at the LHC. Again, our best prediction corresponds to the  cases $\eta/s(T)=0.2$ and $param1$, the rest three cases are shown for testing the sensitivity of this observable to changes of $\eta/s(T)$. Comparing with the ATLAS 2.76 TeV data shown in the figure and Fig.~19 in \cite{Niemi:2015qia}, we can see that only marginal changes in the magnitude of these correlations are expected relative to the lower LHC-energy case, and that the $\eta/s(T)$ discrimination power of the forthcoming new LHC data on this observable is quite similar to that at 2.76 TeV.    

\section{Conclusions}

In this paper, keeping the setup fixed exactly as in Ref.~\cite{Niemi:2015qia}, we have presented the NLO EKRT EbyE model predictions for charged hadron multiplicity and its centrality dependence, for centrality dependence of the flow coefficients $v_n\{2\}$ and correlations of two event-plane angles in the forthcoming $\sqrt{s_{NN}} = 5.023$ TeV Pb+Pb collisions at the LHC. For the 0-5\% centrality class, we predict $dN_{\rm ch}/d\eta\big|_{|\eta|\le 0.5} =1876\dots2046$ and a power-law behavior  $dN_{\rm ch}/d\eta\big|_{|\eta|\le 0.5} \propto s^{0.164\dots0.174}$ from the 200 GeV Au+Au collisions to the 2.76 and 5.023 TeV Pb+Pb collisions. The centrality dependencies of the studied observables are predicted to be very similar to those at 2.76 TeV, and the $v_n\{2\}$ coefficients and event-plane angle correlations are predicted to be almost unchanged. The 5.023 TeV $v_n\{2\}$ measurement may, however, offer slightly more discriminating power on $\eta/s(T)$ than the 2.76 TeV data alone. Especially, the 5.023 TeV LHC multiplicity measurement will be very interesting: already the original EKRT model with ideal fluid dynamics predicted a power law behavior \cite{Eskola:1999fc,Eskola:2001bf} for the cms-energy slope of multiplicity from RHIC to LHC energies but as we discussed here, the cms-energy slope depends particularly on $\eta/s(T)$ of the hottest QGP stage. The forthcoming LHC measurements will give a very welcome increase to the $\sqrt{s_{NN}}$ leverage arm for testing whether the growth of multiplicity indeed continues as a power law in $\sqrt{s_{NN}}$ as we predicted here by keeping the parameters $\beta$ fixed and $K_{\rm sat}$ independent of the cms-energy. 

\textit{Acknowledgements.}
We thank S.S.~R\"as\"anen, K.~Kajantie, and C.~Salgado for discussions, and acknowledge the CSC – IT Center for Science in Espoo, Finland, for the allocation of the computational resources. This work was financially supported by the European Research Council grant HotLHC, No. ERC-2011-StG-279579 (RP), and the Academy of Finland, project 267842 (KT).

\end{document}